\documentclass[reprint,superscriptaddress,floatfix,amsmath,amssymb,pra,aps,showpacs,twocolumn]{revtex4-1}
\usepackage{amsmath,amsfonts,amssymb,amsthm,graphics,graphicx,epsfig,bbm}
\usepackage[colorlinks=true,citecolor=blue,linkcolor=blue,urlcolor=blue]{hyperref}
\usepackage[usenames]{color}
\usepackage{graphicx}
\usepackage{subfigure}
\usepackage{amsmath}
\usepackage{epsfig}
\usepackage{dcolumn}
\usepackage{bm}
\usepackage{color}
\usepackage{times}
\usepackage{epstopdf}
\usepackage{amssymb}
\usepackage{amstext}
\usepackage{latexsym}
\usepackage{hyperref}
\usepackage{amsfonts}
\usepackage{psfrag}
\usepackage{soul,xcolor}
\usepackage[normalem]{ulem}
\usepackage{dsfont}
\usepackage{txfonts}

\newcommand{\ket}[1]{\vert #1 \rangle}

\newcommand{\mean}[1]{\langle #1 \rangle}

\newcommand{\abs}[1]{| #1 |}

\begin{document}
	\setstcolor{red}
	
	\title{Isoenergetic cycle for the quantum Rabi model}% Force line breaks with \\
	
	\author{G. Alvarado Barrios}
	\email[Gabriel Alvarado]{\quad gabriel.alvarado@usach.cl}
	\affiliation{Departamento de F\'isica, Universidad de Santiago de Chile (USACH), 
		Avenida Ecuador 3493, 9170124, Santiago, Chile}
	
	\author{Francisco J. Pe\~na}
	\affiliation{Departamento de F\'isica, Universidad T\'ecnica Federico Santa Mar\'ia Casilla 110V,Valpara\'iso, Chile}
	
	\author{F. Albarr\'an-Arriagada}
	\affiliation{Departamento de F\'isica, Universidad de Santiago de Chile (USACH), 
		Avenida Ecuador 3493, 9170124, Santiago, Chile}
	
	\author{P. Vargas}
	\affiliation{Departamento de F\'isica, Universidad T\'ecnica Federico Santa Mar\'ia Casilla 110V,Valpara\'iso, Chile}
	\affiliation{Center for the Development of Nanoscience and Nanotechnology 9170124, Estaci\'on Central, Santiago, Chile}
	
	\author{J. C. Retamal}
	\affiliation{Departamento de F\'isica, Universidad de Santiago de Chile (USACH), 
		Avenida Ecuador 3493, 9170124, Santiago, Chile}
	\affiliation{Center for the Development of Nanoscience and Nanotechnology 9170124, Estaci\'on Central, Santiago, Chile}	
	\date{\today}% It is always \today, today,
	
	\begin{abstract}
		The isoenergetic cycle is a purely mechanical cycle comprised of adabatic and isoenergetic processes. In the latter the system interacts with an energy bath keeping constant the expectation value of the Hamiltonian. This cycle has been mostly studied in systems consisting of particles confined in a power-law trap. In this work we study the performance of the isoenergetic cycle for a system described by the quantum Rabi model for the case of controlling the coupling strength parameter, the resonator frequency and the two-level system frequency. For the cases of controlling either the coupling strength parameter or the resonator frequency, we find that it is possible to reach maximal unit efficiency when the parameter is sufficiently increased in the first adiabatic stage. In addition, for the first two cases the maximal work extracted is obtained at parameter values corresponding to high efficiency which constitutes an improvement over current proposals of this cycle.
	\end{abstract}
	
	\maketitle
	
	\section{Introduction}
	
	The possibility to create nano-scale devices which are more efficient than current classical counterparts motivates the study of the quantum version of the very well known cycles of classical thermodynamics 
	\cite{Scully2002,Rezek2006Irreversible,Quan2007,Esposito2010,Alvarado2017}. The quantum nature of the working substance and the first law of thermodynamics are the basic ingredients to establish a  relationship between classical thermodynamics and quantum mechanics. 
	
	In the beginning of the 00s a thermodynamical cycle with no classical analogue termed ``isoenergetic cycle" was proposed by Bender \textit{et. al.} \cite{Bender2000}, who envisioned the replacement of the heat baths for so-called ``energy baths". This was originally presented as a  proposal for the substitution of the concept of temperature with the expectation value of the system Hamiltonian \cite{Abe2011,Wang2012,Liu2016,Santos2017}. When the system is coupled to an energy bath it evolves through an isoenergetic process during which the expectation value of the Hamiltonian is constant. This cycle was originally proposed for a non-relativistic single particle confined in one-dimensional potential well. Recently, it was extended to the case of relativistic regime by considering the single-particle Dirac spectrum \cite{Munoz2012,Pena2016} and was treated to the case of an ideal N-particle Fermi system \cite{Wang2015}.
	
	On the other hand, light-matter systems are described in the more basic sense by the quantum Rabi model \cite{Rabi1937}. This model describes the interaction of a single electromagnetic mode with a two-level system (TLS), and it has been studied in a wide range of the coupling parameter \cite{Shore1993,niemczyk2010circuit,Casanova2010DSC}. In particular, the ultrastrong-coupling (USC) regime, which has been experimentally realized  \cite{niemczyk2010circuit}, corresponds to the case where the coupling strength and the resonator frequency become comparable. The light-matter interaction in the USC regime presents interesting properties, such as parity symmetry, and anharmonic energy spectrum \cite{Braak2011}. These properties have led to remarkable applications of systems described by the USC, also termed quantum Rabi systems (QRS), such as fast quantum gates \cite{Romero2012ultrafast}, efficient energy transfer \cite{Kyaw2017,Cardenas-Lopez2017}, and generation of non-classical states \cite{Nori2010,Albarran-Arriagada2017}. Further, current progress in superconducting circuit technology has enabled the manipulation of several parameters of QRSs \cite{Peropadre2010switchable,Gustavsson2012driven,Wallquist2006selective,Sandberg2009exploring,Paauw2009tuning,schwarz2013gradiometric,Koch2007Transmon,Barends2013Xmon,Martinis2014Tunable}. This progress, together with the anharmonic spectrum of the QRS constitutes an interesting system to investigate the performance of the isoenergetic cycle.
	
	In this work we study the isoenergetic cycle where the working substance corresponds to a two-level system interacting with a single electromagnetic mode described by the quantum Rabi model. We consider an analytical approximation of the energy levels which allows for qualitative and quantitative description of the thermodynamical quantities depending on the range of validity of the approximation. We obtain the total work extracted and efficiency of the cycle for the variation of each one of the parameters of the model, namely, the coupling strength, the resonator frequency and the two-level system frequency. For the cases where the energy spectrum shows nonlinearity and degeneracy we see that the cycle performance is improved. In particular, we find that the nonlinear dependence of the energy levels on either the coupling strength, $g$, or the resonator frequency, $\omega$, allows for the cycle efficiency to reach maximal unit value, when the parameter is sufficiently increased in the first adiabatic stage.		
	
	%%%%%%%%%%%%%%%%%%%%%%%%%%%%%%%%%%%%%%%%%%%%%%%%%%%%%%%%%%%%%%%%%%%%%%%%%%%%%%%%%%%%%%
	\subsection{Quantum Rabi model}
	We will consider a working substance composed of a light-matter system described by the quantum Rabi model \cite{Rabi1937,Braak2011}, which reads as	
	%%%%%%%%%%%%%%%%%%%%%%%%%%%%%%%%%%%%%%%%%%%%%%%%%%%%%%%%%%%%%
	\begin{figure}[!ht]
		\centering
		\includegraphics[width=1\linewidth]{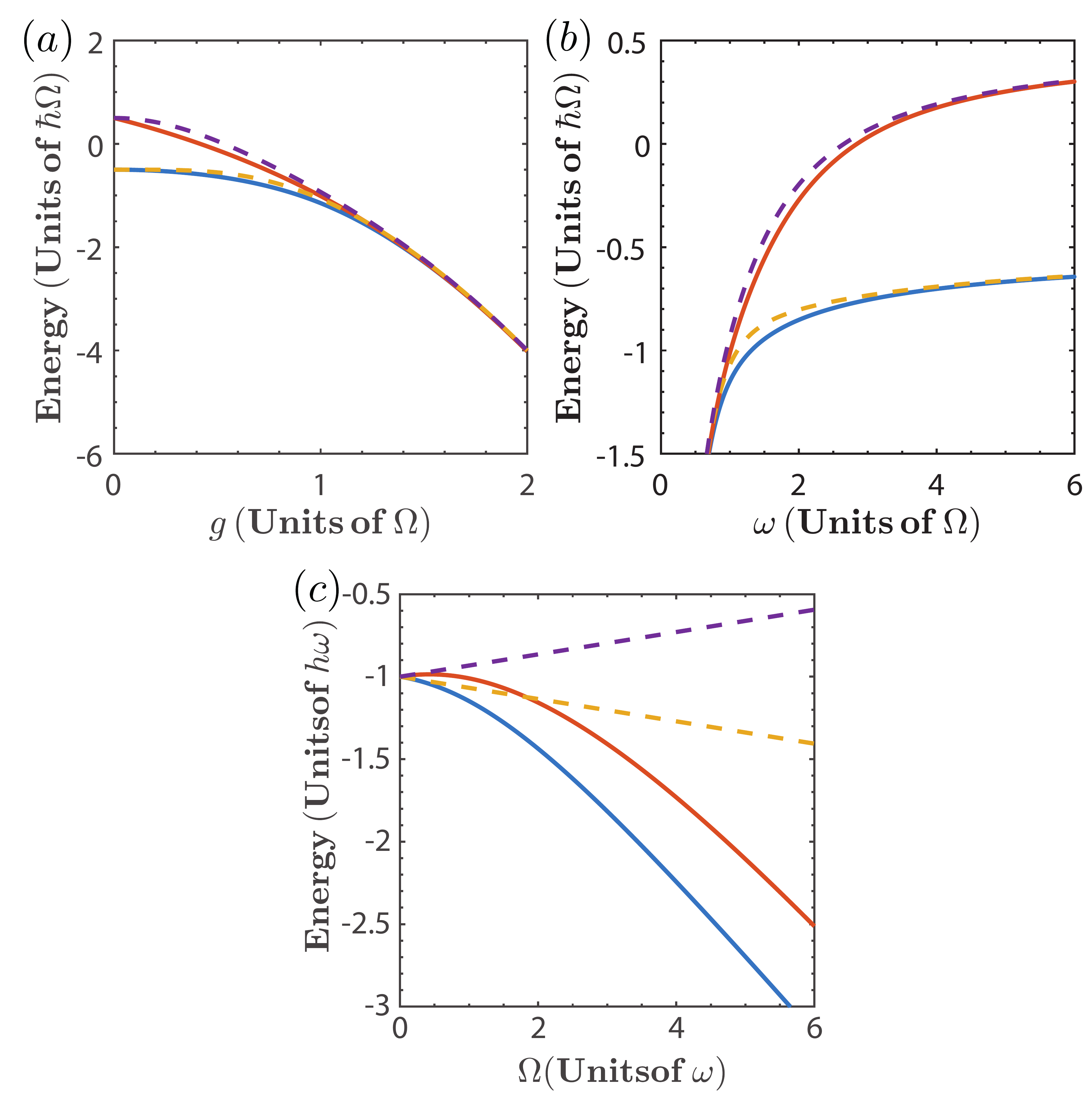}
		\caption{Two lowest energy levels of the quantum Rabi model as a function of (a) the coupling strength $g$, with $\omega=\Omega$, (b) the resonator frequency $\omega$, with $g=\Omega$, and (c) the TLS frequency $\Omega$ with $g=\omega$. The Solid line denotes the exact diagonalization of Eq.~(\ref{Hamiltonian}) and dashed line denotes the approximation given by Eq. (\ref{levels}).}
		\label{Fig1}
	\end{figure}	
	%%%%%%%%%%%%%%%%%%%%%%%%%%%%%%%%%%%%%%%%%%%%%%%%%%%%%%%%%%%%%
	\begin{equation}
	H = \hbar \Omega \sigma_{z} + \hbar \omega a^{\dagger}a + \hbar g \sigma_{x} (a^{\dagger} + a),
	\label{Hamiltonian}
	\end{equation}
	where $a\,(a^{\dagger})$ corresponds to the bosonic annihilation (creation) operator of the resonator mode, $\sigma_{x}$ and $\sigma_{z}$ stand for the Pauli operators describing the two-level system.
	In addition, $\Omega$, $\omega$ and $g$, correspond to the TLS frequency, resonator frequency and TLS-resonator coupling strength, respectively.
	
	This model has been considered for several applications in quantum information processing  \cite{Nataf2011,Romero2012ultrafast,kyaw2015scalable,Kyaw2015QEC,joshi2017qubit,AlbarranArriagada2018}. The ratio between the coupling strength and the resonator frequency $g/\omega$ ($\omega\sim\Omega$) separates the behavior of the system into different regimes \cite{Wolf2013,Rossatto2017}. In the strong coupling regime, where the coupling strength is much larger than any decoherence or dephasing rate in the system, and for values $g/\omega \lesssim 10^{-2}$ one can perform the rotating wave approximation (RWA) and the system can be described by the Jaynes-Cummings model \cite{JC1963}. As the ratio $g/\omega$ is increased beyond the strong coupling regime there is a breakdown of the RWA and the system must be described by the full quantum Rabi model. We distinguish two main regimes for the later case, the ultra-strong coupling regime (USC) \cite{niemczyk2010circuit,FornDiaz2010BSS,Bourassa2009USC} where the coupling strength is comparable to the resonator frequency $g \lesssim \omega$ and the deep-strong coupling regime (DSC) \cite{yoshihara2017DSC,Casanova2010DSC} where the interaction parameter is greater than the relevant frequencies $g > \omega$.
	
	In this work we study the isoenergetic cycle for a working substance which is described by the two lowest energy levels of the quantum Rabi model. In order to better describe the behavior of the thermodynamical figures of merit we will use a simple approximation for the first two lowest energy levels, employed on a recent work \cite{Alvarado2017} based on Refs. \cite{Irish2007,Yu2012analytical}. The approximated energy levels are given by 
	\begin{eqnarray}
	\label{levels}
	E_{\textbf{0}} =&&  - \hbar \frac{g^2}{\omega} - \hbar \frac{\Omega}{2} e^{-2\frac{g^{2}}{\omega^{2}}},\nonumber\\
	E_{\textbf{1}} =&&  - \hbar \frac{g^2}{\omega} + \hbar \frac{\Omega}{2} e^{-2\frac{g^{2}}{\omega^{2}}}, \label{Approximation}
	\end{eqnarray}
	where $E_{\textbf{0}}$ and $E_{\textbf{1}}$ refers to the energy of the ground and first excited state, respectively. Figure \ref{Fig1} shows $E_{\textbf{0}}$ and $E_{\textbf{1}}$ as a function of each of the parameters, $g$, $\omega $ and $\Omega$ as obtained from Eq. (\ref{levels}), compared to their calculation as obtained from the numerical diagonalization of Eq.~(\ref{Hamiltonian}). We can see that the approximation  given by Eq. (\ref{levels}) captures the behavior of the spectrum for all values of $g$ and $\omega$ considered, while for the case of $\Omega$ it is not a good approximation for $\Omega>\omega$. Therefore, we will only consider the numerical calculation for the later case.

	%%%%%%%%%%%%%%%%%%%%%%%%%%%%%%%%%%%%%%%%%%%%%%%%%%%%%%%%%%%%%%%%%%%%%%%%%%%%%%%%%%%%%
	\subsection{\label{sec:firstlaw} First law of thermodynamic}	
	
	Let us consider a system with discrete energy levels and whose Hamiltonian $\hat{H}\left(\xi\right)$ depends explicitly on a parameter $\xi$ that can be varied at an arbitrary slow rate. We define the eigenstate and eigenenergies of $\hat{H}\left(\xi\right)$ by $\hat{H}\vert n;\xi\rangle=E_{n}(\xi)\vert n;\xi\rangle$, then, for state $\ket{\psi}=\sum_{n=0}c_n\ket{\xi;n}$, the average energy $\mean{E}=\mean{\hat{H}}$ of the system takes the form
	\begin{equation}
	\mean{E}=\sum_{n}p_{n}\left(\xi\right)E_{n}\left(\xi\right).
	\end{equation}
	where $p_{n}=\abs{c_{n}}^2$. The change in the average energy due to an arbitrary quasistatic process involving the modulation of the parameter $\xi$ is given by
	\begin{eqnarray}
	\delta\mean{E}=&&\sum_{n}E_{n}\left(\xi\right)\frac{\partial}{\partial\xi}  p_{n}\left(\xi\right)\delta\xi   +\sum_{n}p_{n}\left(\xi\right)\frac{\partial}{\partial\xi}E_{n}\left(\xi\right)\delta\xi \nonumber\\
	=&&\delta Q + \delta W.
	\label{eq1}
	\end{eqnarray}  
	where 
	\begin{eqnarray}
	\delta Q &&= \sum_{n}E_{n}\left(\xi\right)\frac{\partial}{\partial\xi}  p_{n}\left(\xi\right)\delta\xi, \nonumber\\
	\delta W &&= \sum_{n}p_{n}\left(\xi\right)\frac{\partial}{\partial\xi}E_{n}\left(\xi\right)\delta\xi. \label{WorkP}
	\end{eqnarray}
	Equation (\ref{eq1}) is cast in a form reminiscent of the first law of thermodynamics, however, the first term of Eq. (\ref{eq1}) can only be associated with heat when it is possible to define a temperature in the system, as is the case of a interaction with a thermal reservoir in an isochoric process. Since this is not the case for isoenergetic processes, $\delta Q$ is known as the energy exchange \cite{Pena2016,Wang2015}, while the second term $\delta W$ can be identified with the work done. That is, the work done corresponds to the change in the eigenenergies $E_{n}\left(\xi\right)$ which is in agreement with the fact that work can only be performed trough a change in generalized coordinates of the systems, which in turn gives rise to a change in the eigenenergies. 
	
	\subsection{\label{sec:isoenergetic} Isoenergetic Cycle}
	%%%%%%%%%%%%%%%%%%%%%%%%%%%%%%%%%%%%%%%%%%%%%%%%%%%%%%%%%%%%%
	\begin{figure}[t]
		\centering
		\includegraphics[width=1\linewidth]{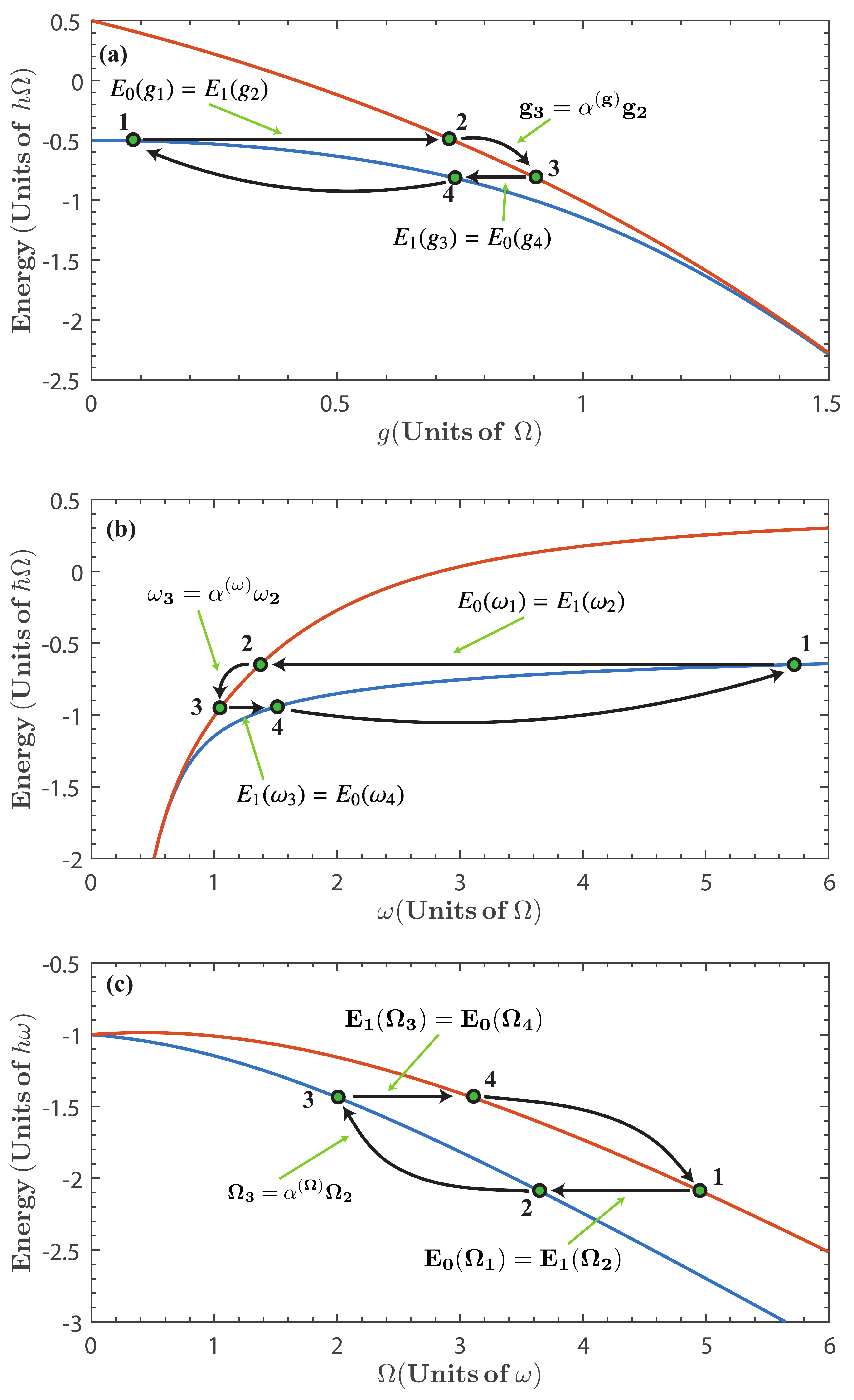}
		\caption{Diagram of the Isoenergetic cycle for (a) $\xi\equiv g$, (b) $\xi\equiv \omega$ and (c) $\xi\equiv \Omega$. Stages $1 \rightarrow 2$ and $3\rightarrow4$ correspond to isoenergetic processes, while stages $2 \rightarrow 3 $ and $4 \rightarrow 1$ correspond to adiabatic processes.}
		\label{Fig2}
	\end{figure}
	%%%%%%%%%%%%%%%%%%%%%%%%%%%%%%%%%%%%%%%%%%%%%%%%%%%%%%%%%%%%%
	The isoenergetic cycle is composed by two adiabatic processes and two isoenergetic ones (see Fig. \ref{Fig2}). In the isoenergetic process the central idea is to keep constant the initial energy expectation value along the procedure, which means $\delta Q + \delta W=0$. Therefore, both work and energy exchange are nonzero during this process. This means that for $\xi$ $\in$ $\left[\xi_{k},\xi_{\ell}\right]$, we have
	\begin{equation}
	\sum_{n}p_{n}(\xi_{k})E_{n}(\xi_{k})=\sum_{n}p_{n}(\xi)E_{n}(\xi)=\sum_{n}p_{n}(\xi_{\ell})E_{n}(\xi_{\ell}),
	\label{eq6}
	\end{equation}
	where $k$ and $\ell$ refers to the ends points of the compression process ($k=1$, $\ell=2$) or expansion process ($k=3$, $\ell=4$). If we consider that the states at the ends of the isoenergetic process correspond to the ground state and first excited state of the system, as is shown in Fig.~\ref{Fig2}, the processes are termed maximal compression for $E_{\textbf{0}}(\xi_{1}) = E_{\textbf{1}}(\xi_{2})$, and maximal expansion for $E_{\textbf{1}}(\xi_{3}) = E_{\textbf{0}}(\xi_{4})$. These conditions yield $\xi_{2}$ as a function of $\xi_{1}$, and $\xi_{4}$ as a function of $\xi_{3}$; and are referred to as the isoenergetic condition.
	
	For a two-level system, the energy exchange along the isoenergetic process for maximal expansion given by \cite{Santos2017,Munoz2012} 
	\begin{eqnarray}
	Q_{\textrm{in}}^{k\rightarrow \ell} =  E_{\textbf{0}}(\xi_{k}) &\times&  \ln\left[\Bigg|\frac{E_{\textbf{0}}(\xi_{\ell})-E_{\textbf{1}}(\xi_{\ell})}{E_{\textbf{0}}(\xi_{k})-E_{\textbf{1}}(\xi_{k})}\Bigg|\right] \nonumber \\ 
	&+& \int_{\xi_{k}}^{\xi_{\ell}}\frac{E_{\textbf{0}}\frac{dE_{\textbf{1}}}{d\xi} - E_{\textbf{1}}\frac{dE_{\textbf{0}}}{d\xi}}{E_{\textbf{0}}(\xi)-E_{\textbf{1}}(\xi)} d\xi. \quad
	\label{energyexchange}
	\end{eqnarray}
	Where $k=1$ and $\ell=2$. For a maximal compression process we refer to the energy exchange as $Q_{\textrm{out}}^{k\rightarrow \ell}$ ($k=3$, $\ell=4$), and it is obtained by exchanging $\textbf{0}$ by $\textbf{1}$, and $\textbf{1}$ by $\textbf{0}$ in Eq. $(\ref{energyexchange})$. The subscripts \textquotedblleft in\textquotedblright $\,$ and \textquotedblleft out\textquotedblright $\,$ denote that energy enters or leaves the system, respectively.
	
	In a isoenergetic process there is work performed the parameter $\xi$ is changed as can be seen from Eq.\ref{WorkP}. At the same time, the energy exchange $Q_{\text{in(out)}}^{k\rightarrow\ell}$ is supplied by the energy bath in order to keep the expectation value of the Hamiltonian constant. Since in this process the average energy change is zero, we write
	\begin{equation}
	Q_{\text{in(out)}}^{k \rightarrow \ell} + W_{\textrm{iso}}^{k\rightarrow \ell}=0.
	\end{equation} 
	Where $W_{\textrm{iso}}^{k\rightarrow\ell}$ is the work done by the system, from this, we obtain that $W_{\textrm{iso}}^{k\rightarrow \ell} = -Q_{\text{in(out)}}^{k \rightarrow \ell}$. As will be seen in the following sections, the isoenergetic processes are the only contribution to the total work extracted. 
	
	On the other hand, in a generic adiabatic process the occupation probabilities $p_{n}(\xi)$ are constant and only work is performed by the system, which is given by \cite{Quan2007}
	\begin{eqnarray}
	\nonumber
	W_{\text{(ad)}}^{i\rightarrow j} &=\int_{\xi_{i}}^{\xi_{j}}d\xi\left(\frac{\partial E}{\partial \xi}\right)_{p_{n}(\xi_{i})=p_{n}(\xi_{j})=\text{constant}} \\ &=\sum_{n}p_{n}(\xi_{i})\left[E_{n}(\xi_{j})-E_{n}(\xi_{i})\right],
	\end{eqnarray}
	where the superscripts $(i, j)$ can taken the values $(i=2,j=3)$ for an adiabatic expansion and $(i=4,j=1)$ for the adiabatic compression, respectively. From Fig.~\ref{Fig2} it is clear that, for each case, the net contribution of the adiabatic processes cancels out, that is, ${W_{\text{(ad)}}^{2 \rightarrow 3} + W_{(ad)}^{4 \rightarrow 1} = 0}$. Therefore, the total work extracted is obtained from the isoenergetic processes, and reads	
	\begin{equation}
	W_{\textrm{total}} = W^{1\rightarrow 2}_{\textrm{iso}} + W^{3\rightarrow 4}_{\textrm{iso}}.
	\end{equation}
	
	Finally, the efficiency of the cycle is
	\begin{equation}
	\eta= \frac{W_{\textrm{total}}}{Q_{\textrm{in}}} = 1 - \frac{Q_{\text{out}}^{3\rightarrow 4}}{Q_{\textrm{in}}^{1\rightarrow 2}}.
	\end{equation}
	It is evident from this expression that to improve the efficiency in a isoenergetic cycle, requires to reduce the ratio $Q^{3\rightarrow 4}_{\textrm{out}}/Q^{1\rightarrow 2}_{\textrm{in}}$. As will be shown later, the quantum Rabi system spectrum yields a better minimization of this ratio than most other systems previously considered.
	
	The isoenergetic cycle is specified by the initial parameter $\xi_{1}$ and $\alpha^{(\xi)} \equiv \xi_{3}/\xi_{2}$, which characterizes the adiabatic process. 
	
	The quantum Rabi model depends of three parameters, the coupling strength $g$, the resonator frequency $\omega$ and the TLS frequency $\Omega$. In our cycle, we will fix two of them and vary the third. Furthermore, we will consider the cases of varying each of the three parameters. 
	
	We have chosen the first isoenergetic process to be of maximal expansion, which will determine whether $\xi$ should be increased or decreased during the first isoenergetic stage. For the case of $\xi = g$ we must increase the parameter, whereas for $\xi = \omega$ and $\xi = \Omega$ we must decrease the parameter. 
	
	%%%%%%%%%%%%%%%%%%%%%%%%%%%%%%%%%%%%%%%%%%%%%%%%%%%%%%%%%%%%%%%%%%%%%%%%%%%%%%%%%%%%%%%%%%%%%%%%%%%%%%%%%%%%
	\section{Quantum Rabi Model as a Working Substance}
	%%%%%%%%%%%%%%%%%%%%%%%%%%%%%%%%%%%%%%%%%%%%%%%%%%%%%%%%%%%%%%%%%%%%%%%%%%%%%%%%%%%%%%%%%%%%%%%%%%%%%%%%%%%%
	\subsection{\label{sec:varying} Case of $\xi\equiv g$}
	\begin{figure}[t]
		\centering
		\includegraphics[width=1\linewidth]{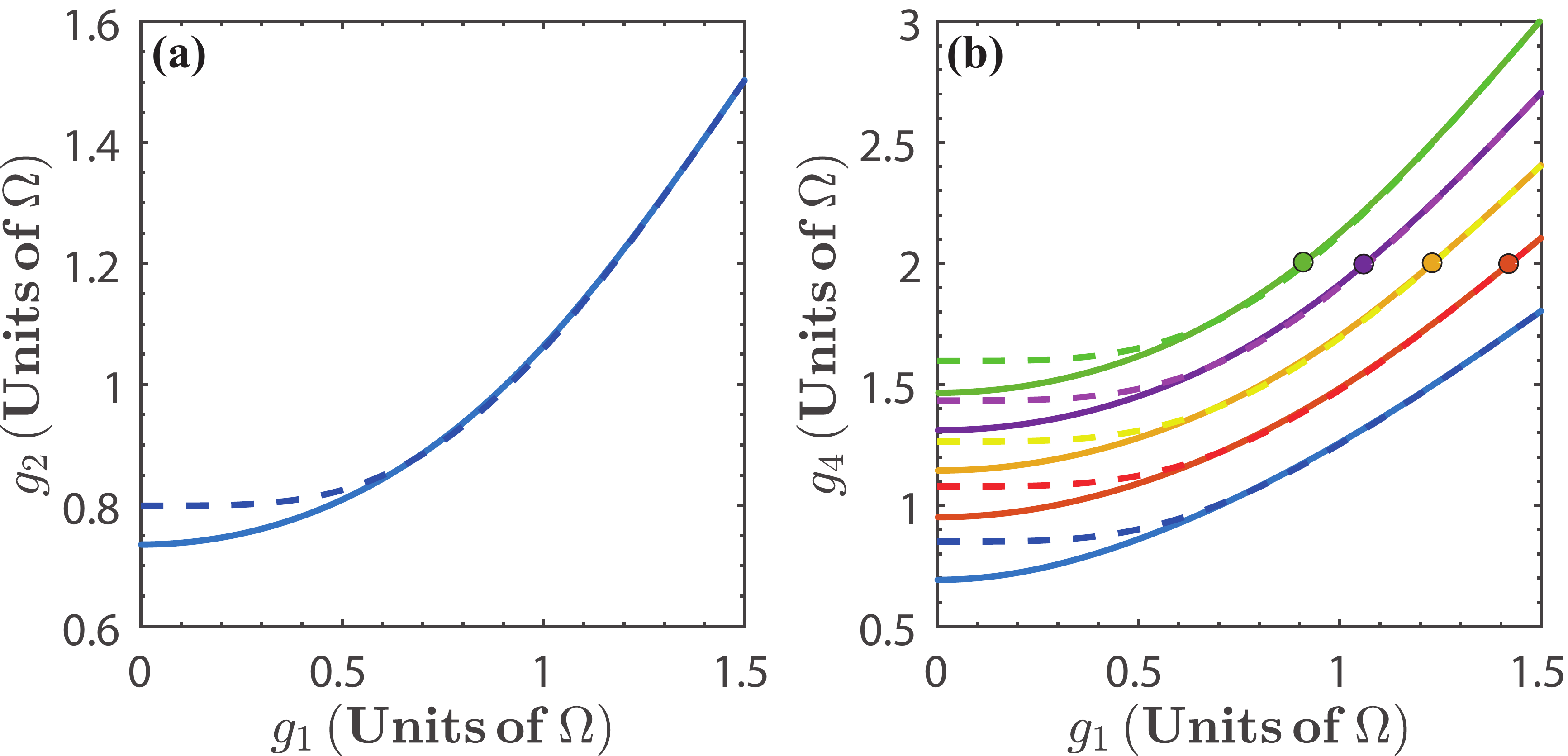}
		\caption{(a) $g_2$ as a function of $g_1$, given by the isoenergetic condition $E_{\textbf{0}}(g_{1})=E_{\textbf{1}}(g_{2})$. (b) $g_4$ as a function of $g_1$, where $g_{4}$ is obtained from the isoenergetic condition $E_{\textbf{1}}(g_{3})=E_{\textbf{0}}(g_{4})$, and $g_{3} = \alpha^{(g)}g_{2}$. We have chosen $\alpha^{(g)}=1.2$ (blue), $\alpha^{(g)}=1.4$ (red), $\alpha^{(g)}=1.6$ (yellow), ${\alpha^{(g)}=1.8}$ (purple) and ${\alpha^{(g)}=2}$ (green). The dots in figure (b) indicate the threshold ${g_{3} = 2 \, \Omega}$ \cite{yoshihara2017DSC}. Solid lines denote the numerical calculation, and dashed lines are calculated with the approximated energy levels.}
		\label{Fig3}
	\end{figure}	
	%%%%%%%%%%%%%%%%%%%%%%%%%%%%%%%%%%%%%%%%%%%%%%%%%%%%%%%%%%%%%%%%%%%%%%%%%	
	Let us start by considering the case of $\xi \equiv g$ as the parameter to be varied, and fix $\omega = \Omega$. This is motivated by experimentally reported control of the coupling strength \cite{Peropadre2010switchable,Gustavsson2012driven}. Figure~\ref{Fig2}~(a) shows the diagram of the isoenergetic cycle corresponding to this case.

	Let us first consider the isoenergetic expansion and compression stages. The first isoenergetic process is subject to the isoenergetic condition given by $E_{\textbf{0}}(g_{1})=E_{\textbf{1}}(g_{2})$ which yields $g_{2}$ as a function of $g_{1}$. This is shown in Fig.~\ref{Fig3}~(a). Due to the structure of the energy levels, the range of values for $g_{1}$ in which the cycle can be operated is approximately between $0<g_{1}<1.5$. Beyond this value, the energy levels become degenerate and we expect  no energy exchange in the isoenergetic process. Therefore, the energy spectrum imposes a bound in the range of values of $g_{1}$ for the operation of the isoenergetic cycle. Similarly we consider the isoenergetic condition for the compression stage $E_{\textbf{1}}(g_{3}) = E_{\textbf{0}}(g_{4})$ and obtain the values of  $g_{4}$ for given $g_{3}$, which is shown in Fig.~\ref{Fig3}~(b).  
	%%%%%%%%%%%%%%%%%%%%%%%%%%%%%%%%%%%%%%%%%%%%%%%%%%%%%%%%%%%%%%%%%%%%%%%%%
	\begin{figure}[t]
		\centering
		\includegraphics[width=1\linewidth]{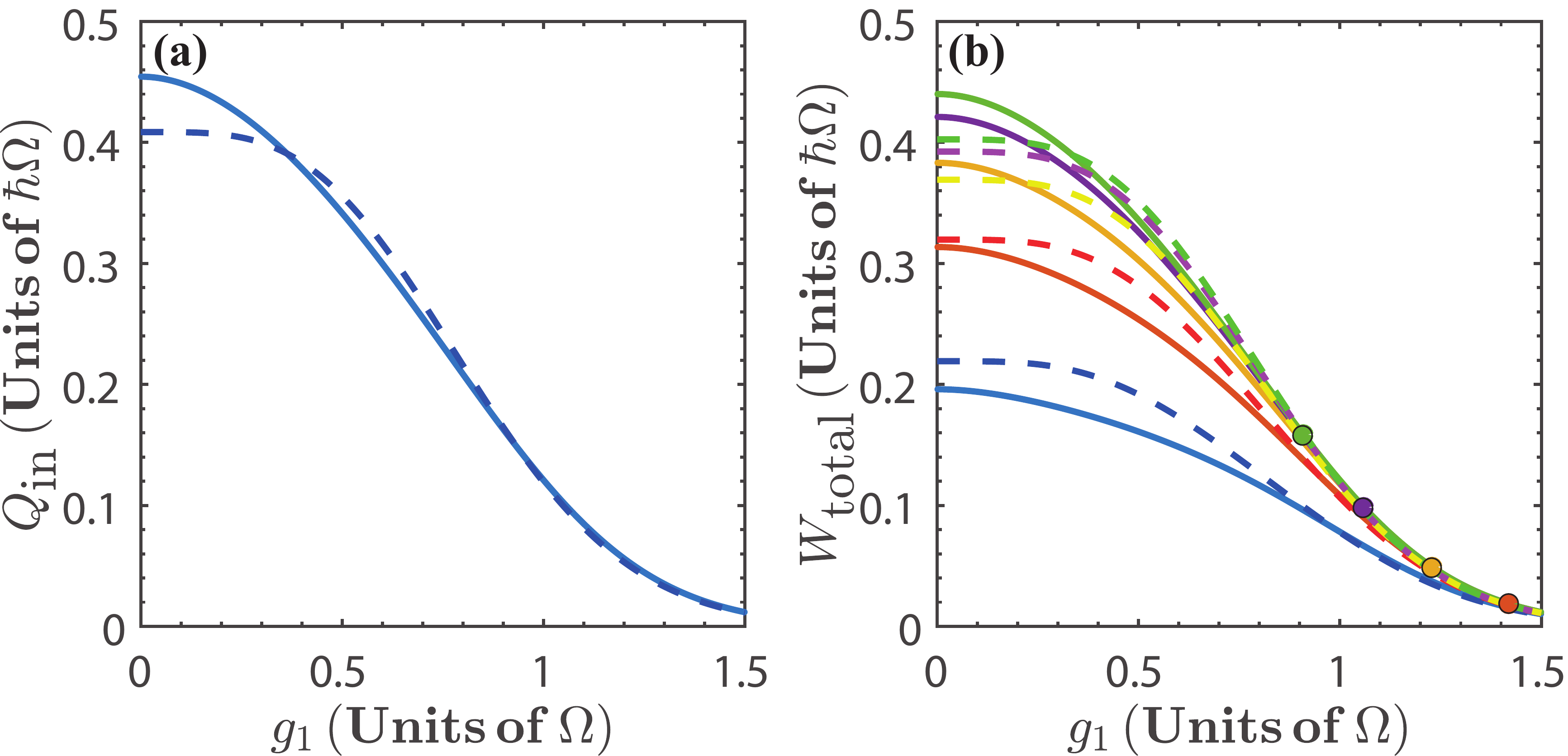}
		\caption{(a) Energy exchange $Q_{\textrm{in}}$ and (b) total work extracted, $W_{\textrm{total}}$, as a function of $g_{1}$, for $\alpha^{(g)}=1.2$ (blue), $\alpha^{(g)}=1.4$ (red), $\alpha^{(g)}=1.6$ (yellow), $\alpha^{(g)}=1.8$ (purple) and $\alpha^{(g)}=2$ (green). The dots in figure (b) indicate the threshold $g_{3} = 2 \, \Omega$ \cite{yoshihara2017DSC}. Solid lines denote the numerical calculation, and dashed lines are calculated with the approximated energy levels.}
		\label{Fig4}      
	\end{figure}	
	%%%%%%%%%%%%%%%%%%%%%%%%%%%%%%%%%%%%%%%%%%%%%%%%%%%%%%%%%%%%%%%%%%%%%%%%%
	\begin{figure}[t]
		\centering
		\includegraphics[width=0.75\linewidth]{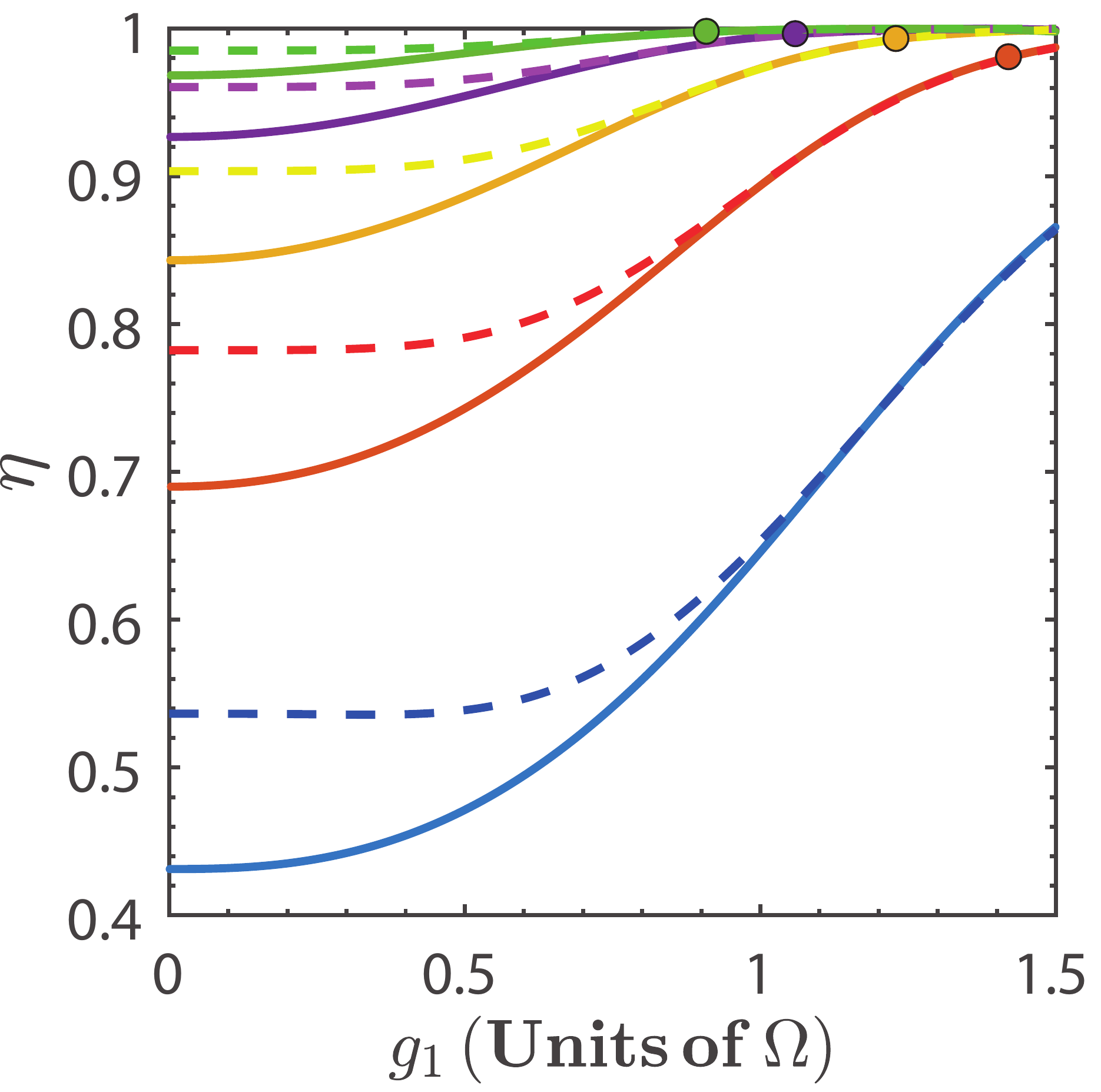}
		\caption{Efficiency $\eta$ as function $g_1$ for $\alpha^{(g)}=1.2$ (blue), $\alpha^{(g)}=1.4$ (red), $\alpha^{(g)}=1.6$ (yellow), $\alpha^{(g)}=1.8$ (purple) and $\alpha^{(g)}=2$ (green). The dots indicate the threshold ${g_{3} = 2 \, \Omega}$ \cite{yoshihara2017DSC}. In both figures solid line denotes the exact numerical calculation, and dashed line is calculated with the approximated energy levels.}
		\label{Fig5}      
	\end{figure}	
	%%%%%%%%%%%%%%%%%%%%%%%%%%%%%%%%%%%%%%%%%%%%%%%%%%%%%%%%%%%%%%%%%%%%%%%%%
	From Eq. (\ref{energyexchange}), we obtain the energy exchange for the isoenergetic expansion and compression process as
	\begin{eqnarray}
	\label{Q12}
	Q^{1\rightarrow 2}_{\textrm{in}}=\frac{2}{\omega^{2}}\left(g_{2}^{2}-g_{1}^{2}\right)\left(\frac{\hbar g_{1}^{2}}{\omega}+\frac{\hbar\Omega}{2}e^{-\frac{2g_{1}^{2}}{\omega^{2}}}\right) \\ \nonumber - \frac{\hbar}{\omega^{3}}\left(\omega^{2}\left(g_{2}^{2}-g_{1}^{2}\right)+\left(g_{2}^{4}-g_{1}^{4}\right)\right),
	\end{eqnarray}
	%%%%%%%%%%%%%%%%%%%%%%%%%%%%%%%%%%%%%%%%%%%%%%%%%%%%%%%%%%%%%
	\begin{eqnarray}
	\label{Q34}
	Q^{3\rightarrow 4}_{\textrm{out}}=\frac{2}{\omega^{2}}\left(g_{3}^{2}-g_{4}^{2}\right)\left(-\frac{\hbar g_{3}^{2}}{\omega}+\frac{\hbar\Omega}{2}e^{-\frac{2g_{3}^{2}}{\omega^{2}}}\right) \\ \nonumber + \frac{\hbar}{\omega^{3}}\left(\omega^{2}\left(g_{3}^{2}-g_{4}^{2}\right)+\left(g_{3}^{4}-g_{4}^{4}\right)\right).
	\end{eqnarray}
	
	We can see from Eq. (\ref{Q12}) and (\ref{Q34}) that the energy exchange that enters or leaves the system is proportional to ${g_2^2-g_1^2}$ or ${g_3^2-g_4^2}$, respectively. Then, by inspecting Fig. \ref{Fig2} (a) we would expect that $Q^{1\rightarrow 2}_{\textrm{in}}/Q^{3\rightarrow 4}_{\textrm{out}}>1$, and that this ratio should be increased by incrementing $\alpha^{(g)}$.
	
	On the other hand, for the first and second adiabatic processes the work done is given by $W^{2\rightarrow 3}=E_{\textbf{1}}(g_{3})-E_{\textbf{1}}(g_{2})$  and $W^{4\rightarrow 1}=E_{\textbf{0}}\left(g_{1}\right)-E_{\textbf{0}}\left(g_{4}\right)$, respectively. Where $g_{3}=\alpha^{(g)} g_{2}$ and $g_{4}$ is specified by the second isoenergetic condition.
	
	The total work extracted, $W_{\textrm{total}}$, depends on $g_{1}$ and $\alpha^{(g)}$, as is shown in Fig. \ref{Fig4} (b). We see from the figure that incrementing the values of the adiabatic parameter $\alpha^{(g)}$ increases the total work extracted, as would be expected. In addition the total work extracted vanishes as $g_{1}\rightarrow 1.5 \Omega$, which is a consequence of the energy levels becoming degenerate at these values of the coupling strength.  
	
	Figure \ref{Fig5} shows the efficiency, $\eta$, of the cycle as a function of $g_{1}$ for different values $\alpha^{(g)}$. From this figure, we see that the efficiency increases with $g_{1}$ as well as with $\alpha^{(g)}$. This is a consequence of the nonlinear dependence of the energy spectrum on the parameter $g$. Additionally, we see that for finite values of $g_{1}$ the efficiency quickly approaches its maximal theoretical value, instead of asymptotically converging to it \cite{Santos2017,Munoz2012,Pena2016}. This can be understood from Fig.~\ref{Fig2}~(a) and Fig.~\ref{Fig3}, since, as $g_{1}$ and $\alpha^{(g)}$ increase, we can expect that the ratio $Q^{3\rightarrow 4}_{\textrm{out}}/Q^{1\rightarrow 2}_{\textrm{in}}$ to be minimized. This is because the nonlinearity of the energy spectrum is such that the second isoenergetic process happen closer to the region where the energy levels become degenerate, and from Eq.~(\ref{Q34}) we see that if $g_{4} \rightarrow g_{3}$, then $Q^{3\rightarrow 4}_{\textrm{out}}\rightarrow 0$. However, this will happen for $W_{\text{total}}\rightarrow 0$ as can be seen from Fig.~\ref{Fig4}~(b). On the other hand, in the region of maximal total work extracted we find values of the efficiency that range in $0.5 <\eta <0.95$ depending on the values of $\alpha^{(g)}$.
	
	%%%%%%%%%%%%%%%%%%%%%%%%%%%%%%%%%%%%%%%%%%%%%%%%%%%%%%%%%%%%%%%%%%%%%%%%%%%%%%%%%%%%%%%%%%%%%%%%%%%%%%
	
	\subsection{\label{sec:varomega} Case of $\xi \equiv \omega$}
	%%%%%%%%%%%%%%%%%%%%%%%%%%%%%%%%%%%%%%%%%%%%%%%%%%%%%%%%%%%%%
	\begin{figure}[t]
		\centering
		\includegraphics[width=1\linewidth]{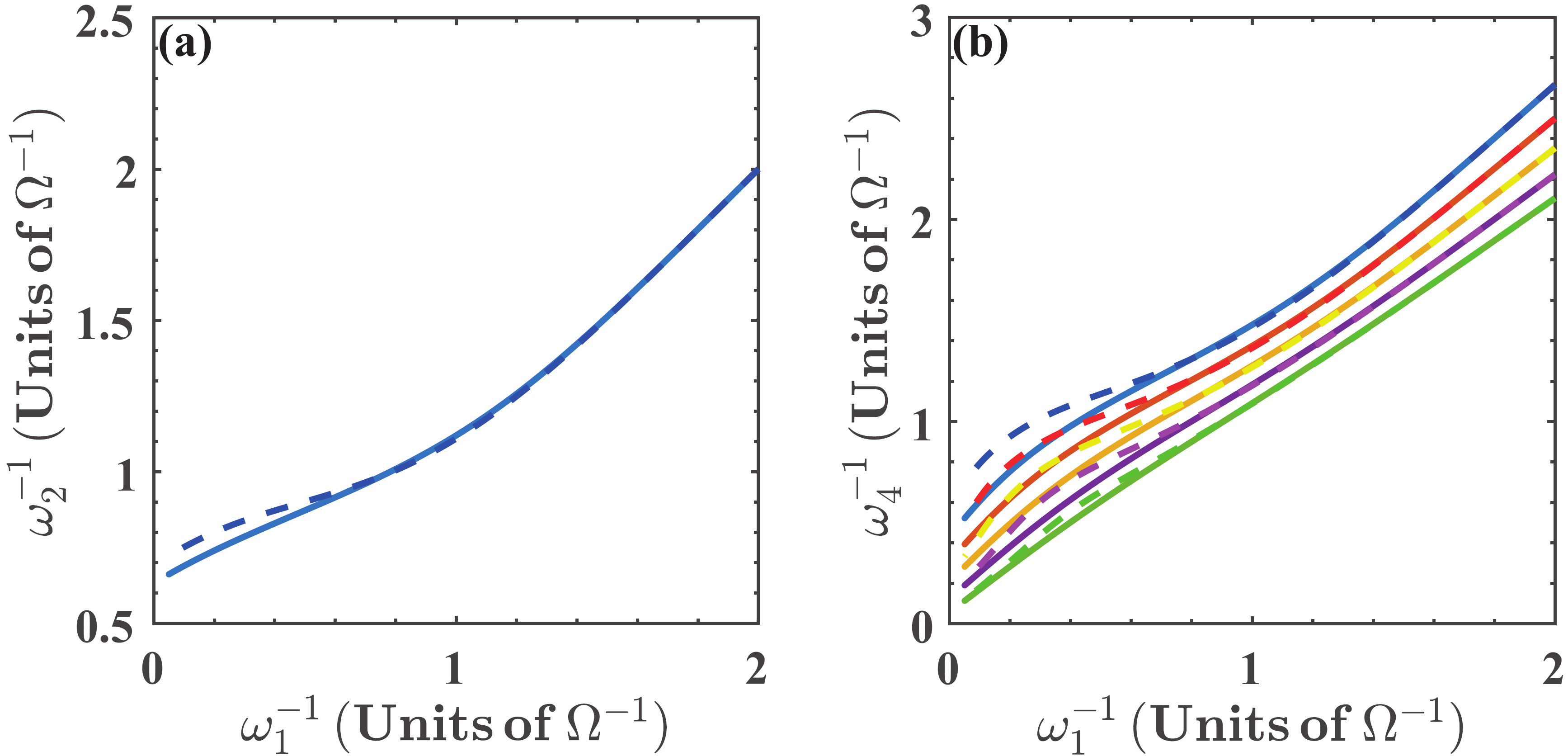}
		\caption{(a) $\omega^{-1}_{2}$ as a function of $\omega^{-1}_{1}$ given by the isoenergetic condition $E_{\textbf{0}}(\omega_{1})=E_{\textbf{1}}(\omega_{2})$. (b) $\omega_{4}^{-1}$ as a function of $\omega_{1}^{-1}$ where $\omega_{4}$ is obtained from the isoenergetic condition $E_{\textbf{1}}(\omega_{3})=E_{\textbf{0}}(\omega_{4})$, and $\omega_{3} = \alpha^{(\omega)}\omega_{2}$. We have chosen $\alpha^{(\omega)}=0.75$ (blue), $\alpha^{(\omega)}=0.80$ (red), $\alpha^{(\omega)}=0.85$ (yellow), $\alpha^{(\omega)}=0.90$ (purple) and $\alpha^{(\omega)}=0.95$ (green). Solid lines denote the numerical calculation, and dashed lines are calculated with the approximated energy levels.}
		\label{Fig6}      
	\end{figure}	
	%%%%%%%%%%%%%%%%%%%%%%%%%%%%%%%%%%%%%%%%%%%%%%%%%%%%%%%%%%%%%
	
	Now, we consider the choice of $\xi \equiv \omega$ as the parameter to be varied, and fix $g = \Omega$. This is motivated by experimentally reported control of the resonator frequency \cite{Wallquist2006selective,Sandberg2009exploring}.
	
	For this case, the energy exchange for maximal expansion and compression are given by	
	\begin{eqnarray}
	Q^{1 \rightarrow 2}_{\textrm{in}} &=& \left(-2g^{2}\left(\frac{1} {\omega_{2}^{2}}-\frac{1}{\omega_{1}^{2}}\right)\right) E_{1}\left(\omega_{1}\right)\\ \nonumber
	&-& \frac{4}{3}\hbar g^{4}\left(\frac{1}{\omega_{2}^{3}}-\frac{1}{\omega_{1}^{3}}\right) -\hbar g^{2}\left(\frac{1}{\omega_{2}}-\frac{1}{\omega_{1}}\right),
	\end{eqnarray}
	\begin{eqnarray}
	Q^{3\rightarrow 4}_{\textrm{out}}&=& \left(2g^{2}\left(\frac{1} {\omega_{3}^{2}}-\frac{1}{\omega_{4}^{2}}\right)\right)  E_{2}\left(\omega_{3}\right)\\ \nonumber
	&+& \frac{4}{3}\hbar g^{4}\left(\frac{1}{\omega_{3}^{3}}-\frac{1}{\omega_{4}^{3}}\right) +\hbar g^{2}\left(\frac{1}{\omega_{3}}-\frac{1}{\omega_{4}}\right).
	\end{eqnarray}	
	Where $\omega_{2}$, and $\omega_{4}$ are obtained from the isoenergetic conditions $E_{\textbf{1}}(\omega_{2}) = E_{\textbf{0}}(\omega_{1})$ and  $E_{\textbf{0}}(\omega_{4}) = E_{\textbf{1}}(\omega_{3})$, respectively. This is presented in Fig.~\ref{Fig6}. In what follows we find convenient to express the results in terms of $1/\omega_{1}$.
	%%%%%%%%%%%%%%%%%%%%%%%%%%%%%%%%%%%%%%%%%%%%%%%%%%%%%%%%%%%%%
	\begin{figure}[t]
		\centering
		\includegraphics[width=1\linewidth]{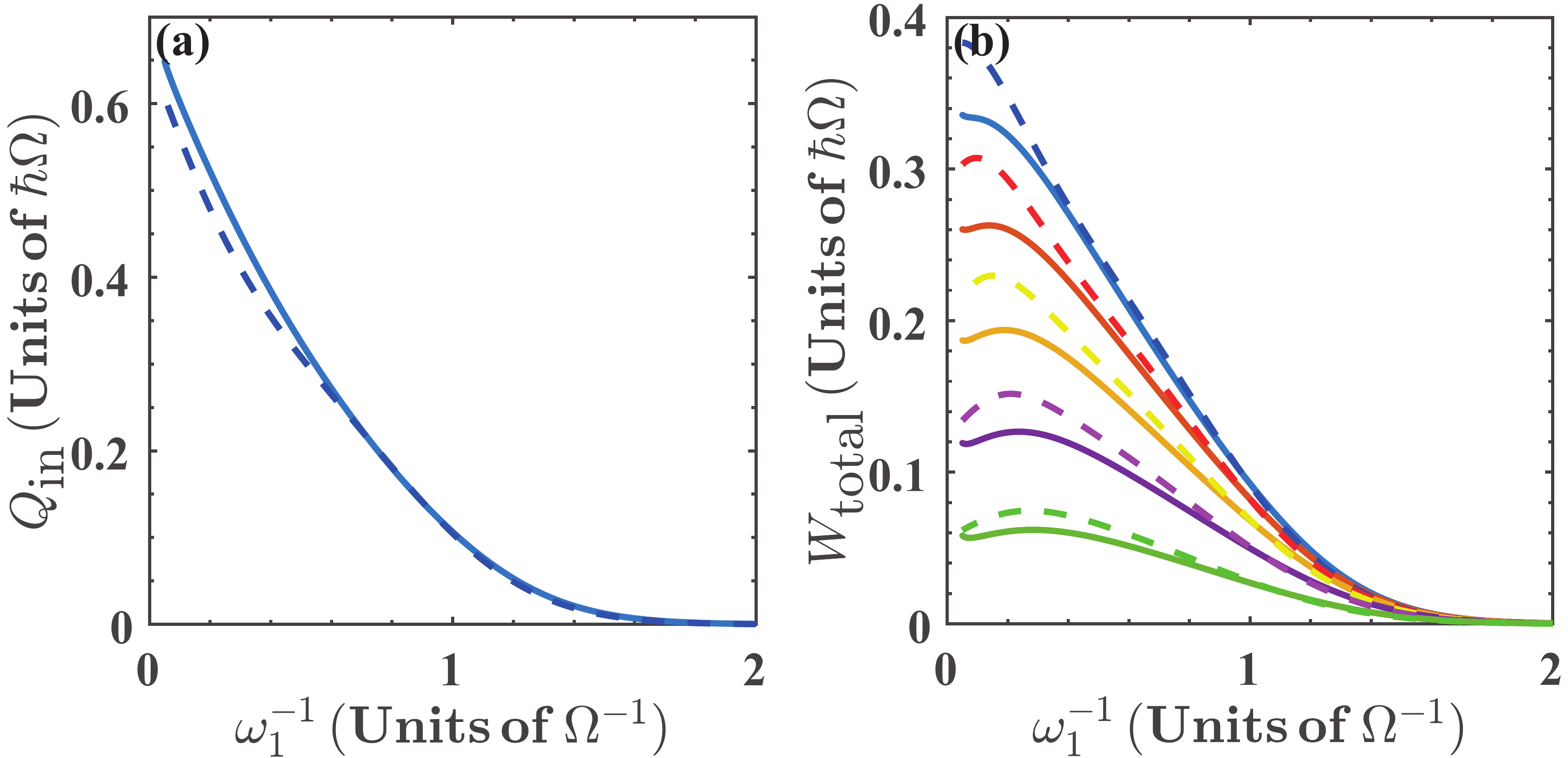}
		\caption{(a) Energy exchange $Q_{\textrm{in}}$ and (b) total work extracted $(W_{\textrm{total}})$ as a function of $\omega_{1}^{-1}$ for $\alpha^{(\omega)}=0.75$ (blue), $\alpha^{(\omega)}=0.8$ (red), $\alpha^{(\omega)}=0.85$ (yellow), $\alpha^{(\omega)}=0.90$ (purple) and $\alpha^{(\omega)}=0.95$ (green). Solid lines denote the numerical calculation, and dashed lines are calculated with the approximated energy levels.}
		\label{Fig7}      
	\end{figure}	
	%%%%%%%%%%%%%%%%%%%%%%%%%%%%%%%%%%%%%%%%%%%%%%%%%%%%%%%%%%%%%	
	\begin{figure}[t]
		\centering
		\includegraphics[width=0.75\linewidth]{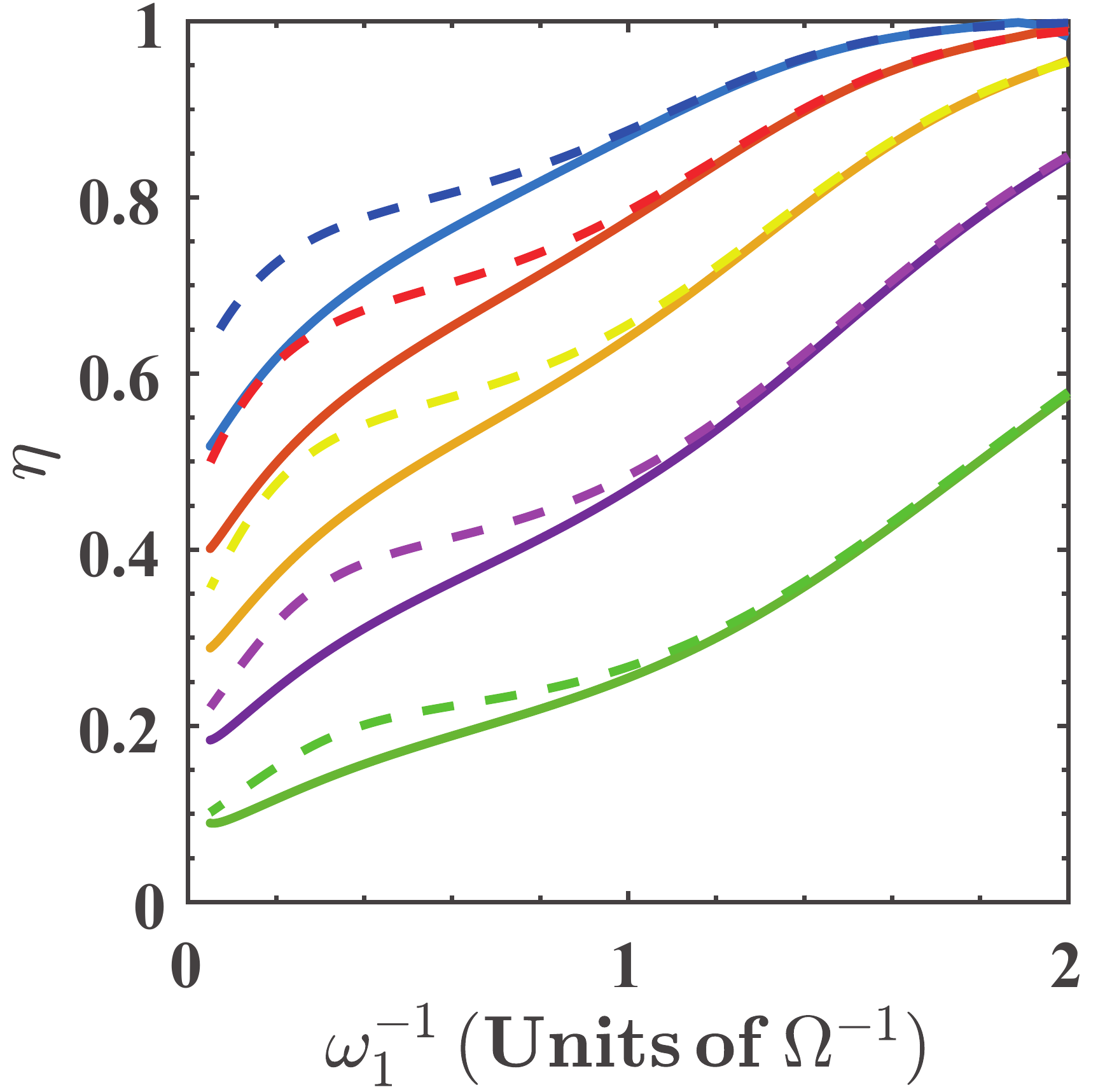}
		\caption{Efficiency as function $\omega_{1}^{-1}$ for $\alpha^{(\omega)}=0.75$ (blue), $\alpha^{(\omega)}=0.8$ (red), $\alpha^{(\omega)}=0.85$ (yellow), $\alpha^{(\omega)}=0.90$ (purple) and $\alpha^{(\omega)}=0.95$ (green). Solid lines denote the numerical calculation, and dashed lines are calculated with the approximated energy levels.}
		\label{Fig8}      
	\end{figure}
	%%%%%%%%%%%%%%%%%%%%%%%%%%%%%%%%%%%%%%%%%%%%%%%%%%%%%%%%%%%%%
	
	In this case, the range of values of $\omega$ for the operation of the isoenergetic cycle is lower bounded by $\omega = 0.5 \, \Omega$. Below this value the energy levels become degenerate and there is no total work extracted nor energy exchange as can be seen from Fig. \ref{Fig7}. 
	
	For the first and second adiabatic processes the work done is given by $W^{2\rightarrow 3} = E_{\textbf{1}}\left(\omega_{3}\right)-E_{\textbf{1}}\left(\omega_{2}\right)$ and ${W^{4\rightarrow 1} = E_{\textbf{0}}\left(\omega_{1}\right)-E_{\textbf{0}}\left(\omega_{4}\right)}$, respectively. Where ${\omega_{3}=\alpha^{(\omega)}\omega_{2}}$, and $\omega_{4}$ is specified by the second isoenergetic process.
	
	The total work extracted, $W_{\textrm{total}}$, is shown in Fig.~\ref{Fig7}~(a) as a function of $\omega_{1}^{-1}$. We see that for ${0.35 \lesssim \omega_{1}^{-1} \lesssim 0.45}$ (in units of $\Omega^{-1}$) we obtain the region of maximal $W_{\textrm{total}}$ for different values of $\alpha^{(\omega)}$. In addition, in Fig.~\ref{Fig7} we see that as $\omega_{1}\rightarrow 2 \, \Omega^{-1}$, then, $Q_{\text{in}}^{1\rightarrow 2}\rightarrow 0$ and $W_{\text{total}}\rightarrow 0$, which is a consequence of the energy levels becoming degenerate beyond this value of resonator frequency.
	%%%%%%%%%%%%%%%%%%%%%%%%%%%%%%%%%%%%%%%%%%%%%%%%%%%%%%%%%%%%%
	\begin{figure}
		\includegraphics[width=\linewidth]{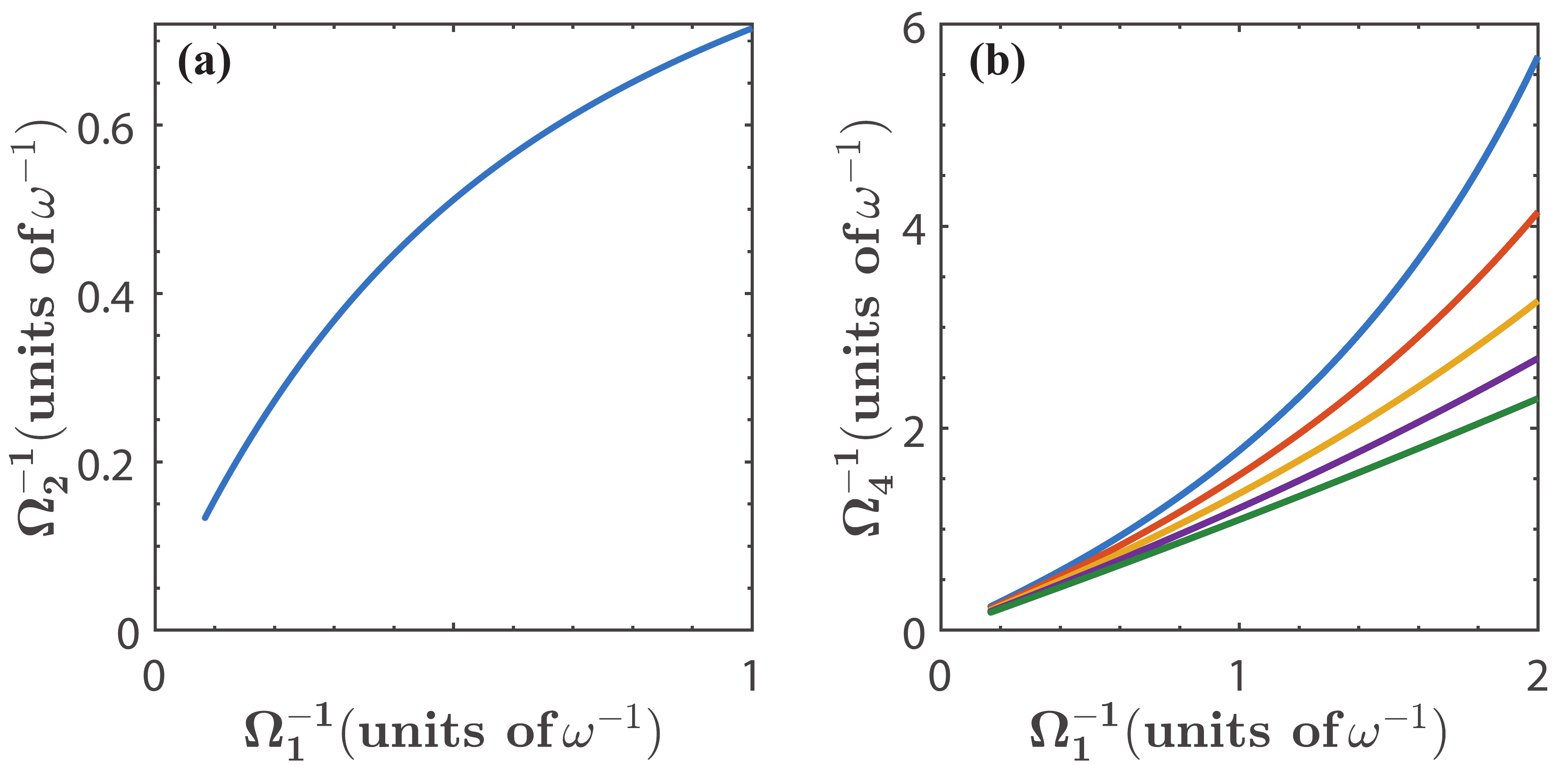}
		\caption{(a) Shows $\Omega_{2}^{-1}$ as a function of $\Omega_{1}^{-1}$ given by the isoenergetic condition $E_{\textbf{0}}(\Omega_{1})=E_{\textbf{1}}(\Omega_{2})$. (b) Shows $\Omega_{4}^{-1}$ as a function of $\Omega_{1}^{-1}$ where $\Omega_{4}$ is obtained from the isoenergetic condition $E_{\textbf{1}}(\Omega_{3})=E_{\textbf{0}}(\Omega_{4})$, and $\Omega_{3} = \alpha^{(\Omega)}\Omega_{2}$. We have chosen $\alpha^{(\Omega)}=0.75$ (blue), $\alpha^{(\Omega)}=0.8$ (red), $\alpha^{(\Omega)}=0.85$ (yellow), $\alpha^{(\Omega)}=0.90$ (purple) and $\alpha^{(\Omega)}=0.95$ (green). In this case we have only considered the exact numerical calculation.}
		\label{IsoCond_WQ}
	\end{figure}
	%%%%%%%%%%%%%%%%%%%%%%%%%%%%%%%%%%%%%%%%%%%%%%%%%%%%%%%%%%%%%
	
	In Fig. \ref{Fig8} we show the efficiency as a function of $\omega_{1}^{-1}$ for different values of $\alpha^{(\omega)}$, where we see that the efficiency increases as $\alpha^{(\omega)}$ is reduced. Notice that the efficiency approaches its maximal theoretical value within the range of $\omega_{1}$ considered. The reason for this is similar to the case of $\xi=g$, where the nonlinearity and degeneracy of the energy spectrum lead to a minimization of the ratio $Q^{3\rightarrow 4}_{\textrm{out}}/Q^{1\rightarrow 2}_{\textrm{in}}$. This can be seen in Fig.~\ref{Fig2}~(b). At the same time, the maximization of the efficiency occurs as the energy exchange and total work extracted go to zero. On the other hand, in the region of maximal total work extracted we find values of the efficiency that range in $0.1 <\eta <0.65$ depending on the values of $\alpha^{(\omega)}$.

	For both the $\xi\equiv g$ case or the $\xi\equiv\omega$ case, the nonlinearity and degeneracy of the energy spectrum allows to reach maximal efficiency of the isoenergetic cycle.
	
	\subsection{Case of $\xi \equiv \Omega$}
	For the final case, we consider the choice $\xi \equiv \Omega$ as the parameter to be varied, and fix $g=\omega$. This is motivated by experimentally reported control of the TLS frequency \cite{Paauw2009tuning,schwarz2013gradiometric}. Since the approximation of Eq.(\ref{Approximation}) breaks down for $\Omega>\omega$, we will only consider numerical calculations of the figures of merit. 
	
	The solution for the isoenergetic condition is shown in Fig. (\ref{IsoCond_WQ}). We see that this case differs from the previous ones in that there is no need to limit the parameter $\Omega$ to a specific range of values  because there is no degeneracy of the energy levels. Nonetheless, we have restricted the values of $\Omega$ to the range $0.5<\Omega<6$ (in units of $\omega$) to facilitate the comparison with the other cases.
	
\begin{figure}[t!]
		\includegraphics[width=\linewidth]{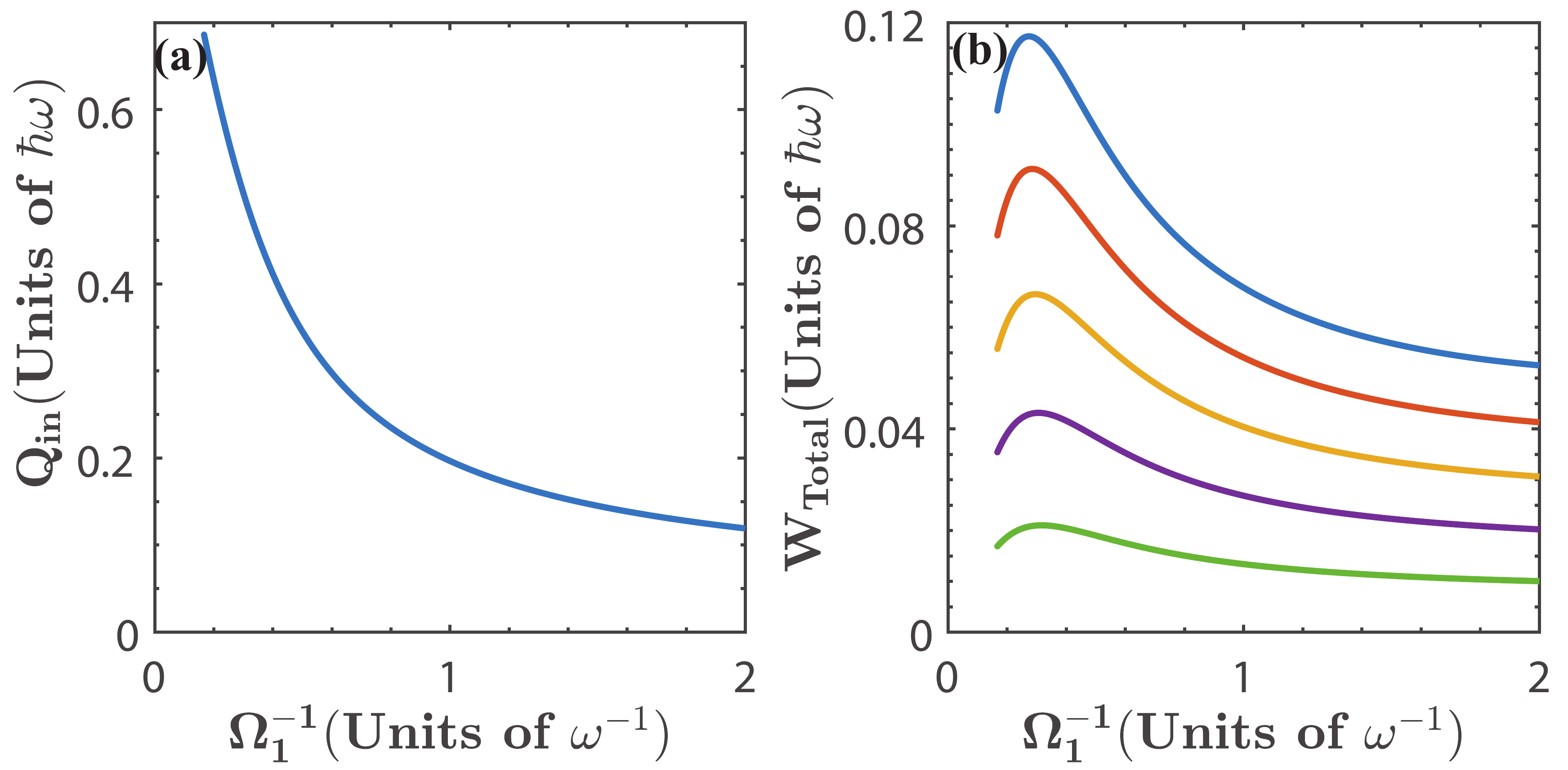}
		\caption{(a) Energy exchange $Q_{\textrm{in}}$ and (b) total work extracted $(W_{\textrm{total}})$ as a function of $\Omega^{-1}_{1}$ for $\alpha^{(\Omega)}=0.75$ (blue), $\alpha^{(\Omega)}=0.8$ (red), $\alpha^{(\Omega)}=0.85$ (yellow), $\alpha^{(\Omega)}=0.90$ (purple) and $\alpha^{(\Omega)}=0.95$ (green).}
		\label{Work_WQ}
	\end{figure}

	The total work extracted is shown in Fig.~(\ref{Work_WQ}), it can be seen that it is considerably smaller than in previous cases, as expected from inspecting the energy spectrum in Fig.~(\ref{Fig2})~(c). Since in this case there is no degeneracy, the total work extracted does not vanish within the chosen range of the parameter.
	
\begin{figure}[t!]
		\includegraphics[width=0.75\linewidth]{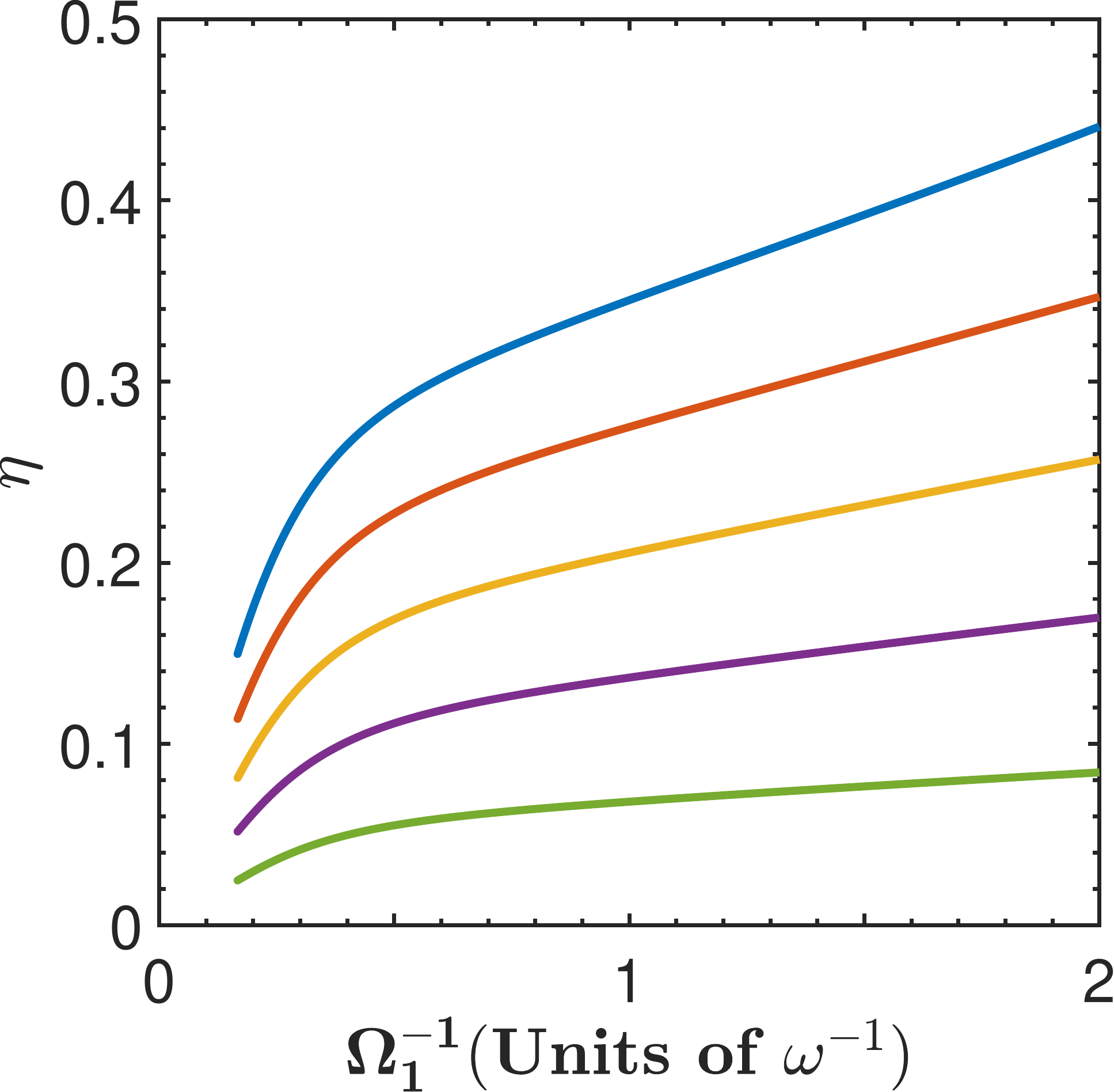}
		\caption{Efficiency as a function of $\Omega_{1}^{-1}$ for different values of $\alpha^{(\Omega)}$ given by $\alpha^{(\Omega)}=0.75$ (blue), $\alpha^{(\Omega)}=0.8$ (red), $\alpha^{(\Omega)}=0.85$ (yellow), ${\alpha^{(\Omega)}=0.90}$ (purple) and $\alpha^{(\Omega)}=0.95$ (green).}
		\label{Eff_WQ}
	\end{figure}

	In Fig.~(\ref{Eff_WQ}) we show the efficiency as a function of $\Omega_{1}^{-1}$ for different values of $\alpha^{(\Omega)}$. Here, the efficiency is smaller than in the previous cases. This is because the functional dependence of the energy levels with $\Omega$ is closer to linear behavior as compared with the other two parameters that were previously considered. 
	%%%%%%%%%%%%%%%%%%%%%%%%%%%%%%%%%%%%%%%%%%%%%%%%%%%%%%%%%%%%%
	%%%%%%%%%%%%%%%%%%%%%%%%%%%%%%%%%%%%%%%%%%%%%%%%%%%%%%%%%%%%%
	%%%%%%%%%%%%%%%%%%%%%%%%%%%%%%%%%%%%%%%%%%%%%%%%%%%%%%%%%%%%%
	\section{Conclusions}
	We have studied the performance of an isoenergetic cycle with a working substance described by the quantum Rabi model. We have considered the variation of each of the parameters of the system, $g$, $\omega$ and $\Omega$. We use a simple approximation of the energy levels which helps to understand the behavior of the figures of merit.
	
	We find that the nonlinear dependence of the energy levels on either the coupling strength, $g$, or the resonator frequency, $\omega$, allows for the cycle efficiency to reach maximal unit value. This occurs when the parameter is sufficiently increased (for $g$) or decreased (for $\omega$) in the first adiabatic stage. On the other hand, maximal total work extracted is found at efficiencies in the range of $0.5<\eta<0.95$ for the variation of $g$, and in the range of $0.1< \eta <0.65$ for the variation of $\omega$, which depend on the changes induced by the adiabatic processes. 
	
	Finally, we considered the case of varying the TLS frequency $\Omega$. We find that the total work extracted and the efficiency are considerably smaller than in the previous cases. This is because the functional dependence of the energy levels with $\Omega$ is closer to linear behavior as compared with the other two parameters.
	
	Therefore, we can distinguish the degeneracy and nonlinearity of the energy spectrum of the working substance as optimal conditions to consider for the isoenergetic cycle. 
	
	The authors acknowledge support from CONICYT Doctorado Nacional 21140587, CONICYT Doctorado Nacional 21140432, Direcci\'on de Postgrado USACH, FONDECYT-postdoctoral 3170010, Financiamiento Basal para Centros Cient\'ificos y Tecnol\'ogicos de Excelencia, under Project No. FB 0807 (Chile), USM-DGIIP grant number PI-M-17-3 (Chile) and FONDECYT under grant No. 1140194.

\end{document}